\documentclass[draft]{aipproc}
\usepackage{psfig}

\layoutstyle{6x9}

\def\kms{km~s$^{-1}$}
\def\farcs{\hbox{$.\!\!^{\prime\prime}$}}

\begin{document}

\title{Fifteen Years of High-Resolution Radio Imaging of Supernova 1987A}


\author{B. M. Gaensler}{
  address={School of Physics, The University of Sydney, Sydney NSW,
Australia},
altaddress={ARC Federation Fellow},
%altaddress={Harvard-Smithonian Center for Astrophysics, Cambridge MA, USA}
}

\author{L. Staveley-Smith}{
  address={School of Physics, The University of Western Australia,
Crawley WA, Australia},
altaddress={Premier's Fellow in Radio Astronomy}
}

\author{R. N. Manchester$^\dagger$, M.~J.~Kesteven, L. Ball and A. K. Tzioumis}{
  address={Australia Telescope National Facility, CSIRO, Marsfield NSW,
Australia}
}

\begin{abstract}

Supernova 1987A in the Large Magellanic Cloud provides a spectacularly
detailed view of the aftermath of a core-collapse explosion.  The
supernova ejecta initially coasted outward at more than 10\% of the
speed of light, but in 1990 were observed to decelerate rapidly as
they began to encounter dense circumstellar material expelled by
the progenitor star. The resulting shock has subsequently produced
steadily brightening radio synchrotron emission, which is resolved
by  the Australia Telescope Compact Array (ATCA) into an expanding
limb-brightened shell.  Here we present 15 years of ATCA imaging
of Supernova~1987A, at an effective angular resolution of $0\farcs4$.
We find that the radio remnant has accelerated in its expansion
over this period, from $\approx3600$~\kms\ in 1992 to $\approx5200$~\kms\
at the end of 2006. The published diameters of the evolving X-ray
shell have been $\sim15\%$ smaller than the corresponding radio
values, but a simultaneous Fourier analysis of both radio and X-ray
data eliminates this discrepancy, and yields a current diameter for
the shell in both wave-bands of $\approx1\farcs7$.  An asymmetric
brightness distribution is seen in radio images at all ATCA epochs:
the eastern and western rims have higher fluxes than the northern
and southern regions, indicating that most of the radio emission
comes from the equatorial plane of the system, where the progenitor
star's circumstellar wind is thought to be densest.  The eastern
lobe is brighter than and further from the supernova site than the
western lobe, suggesting an additional asymmetry in the initial
distribution of supernova ejecta.

\end{abstract}

\maketitle

%%%%%%%%%%%%%%%%%%%%%%%%%%%%%%%%%%%%%%%%%%%%
%% MAINMATTER
%%%%%%%%%%%%%%%%%%%%%%%%%%%%%%%%%%%%%%%%%%%%

\section{Introduction}

\begin{figure}[t!]
%\begin{wrapfigure}{r}{0.5\textwidth}
\centerline{\psfig{file=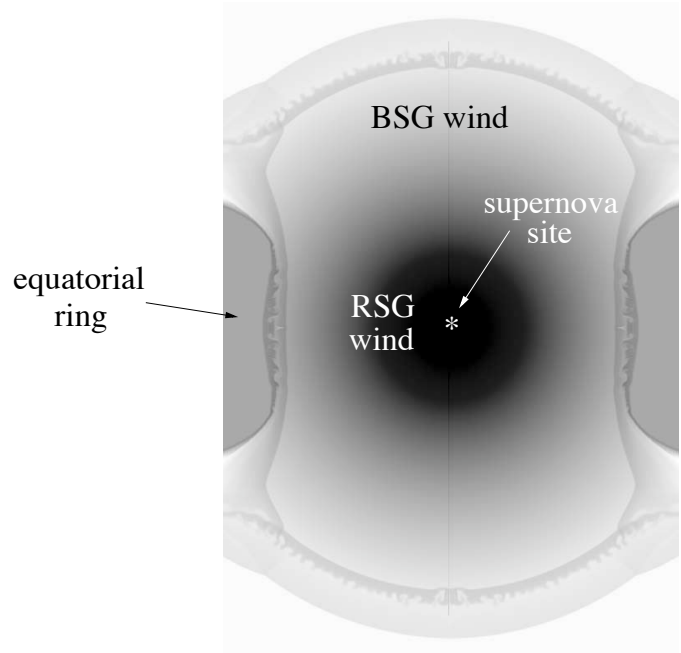,width=0.6\textwidth}}
\vspace{-0.15cm}
%{\footnotesize {\bf FIGURE 1.}
\caption{
Hydrodynamic simulation of the CSM around SN~1987A, adapted
from \cite{mmc+03}. The greyscale represents density, and the system
is viewed from the side, such that the equatorial ring
is seen edge-on and the nebular symmetry axis is oriented vertically.}
\label{fig_hydro}
%\end{wrapfigure}
\end{figure}

Supernova (SN) 1987A continues to be an amazing laboratory for
studying the interaction between the circumstellar medium (CSM) of
a massive star and the subsequent SN ejecta. While the nature of
the progenitor star of SN~1987A is still not completely
clear \cite{mp07,smi07}, a
reasonable consensus 
is that late in its life, Sk~--69$^\circ$202 was a red
supergiant (RSG), and produced a dense, slow, wind focused into the
equatorial plane of the system. About 20\,000 years before
core-collapse, the star evolved into a blue supergiant (BSG), and
began to produce a high-velocity, low density, isotropic wind. The
interaction between the BSG and RSG stellar winds, combined with the
photo-ionization of the RSG wind by UV photons produced by the BSG,
produced the optical ``triple-ring'' structure seen after the SN
(e.g., \cite{bkh+95}).  The subsequent radio emission produced by
the interaction of the expanding ejecta with the swept-up CSM needs
to be interpreted within the context of this complex pre-existing
set of structures (see Fig.~\ref{fig_hydro}).

An initial burst of radio emission from SN~1987A was seen by the
Molonglo Observatory Synthesis Telescope (MOST) at an observing
frequency of 843~MHz, just two days after core-collapse \cite[]{tcb+87}.
At this very early stage the radio light-curve was still rising;
the flux peaked on day 4, and then exhibited a power-law decay,
fading below the detection threshold by day 150. This prompt phase
of emission has been interpreted as synchrotron emission produced
as the shock moved through the inner regions of the CSM, the swift
decay of the flux resulting from rapid motion of the SN shock through
the $\rho \propto r^{-2}$ radial density profile of the BSG stellar
wind \cite[]{sm87,cf87}.  At these early epochs, H$\alpha$ and radio
VLBI observations both indicate that the shock velocity was $\approx
19\,000 - 30\,000$~\kms\ \cite[]{hd87,jkb+88}.

Diligent radio monitoring over subsequent years was
rewarded in mid-1990, when the MOST redetected radio emission
from SN~1987A.  This
was quickly followed by confirmation at higher frequencies with the
Australia Telescope Compact Array (ATCA).
At 9~GHz (the ATCA's highest observing frequency at
that time), a diffraction-limited angular resolution of $\approx0\farcs9$
could be reached. Such data demonstrated that the new radio
source was spatially extended, presumably tracing the interaction between the
expanding shock and the swept-up CSM \cite[]{smk+92,sbr+93}.  Here
we provide an update on our on-going ATCA imaging campaign (see
previous reports by \cite{gms+97,mgw+02,mgs+05}), and discuss what
we have have learned from this unique view of an evolving young
supernova remnant.  The corresponding evolution of the flux density
and spectral index are presented by \cite{sta07} (see also 
\cite{mgw+02,bch+01}, and references therein).

\section{Expansion of the Supernova Remnant}

Since 1992, a deep 9-GHz ATCA observation of SN~1987A  has been
carried out approximately every six months. Over most of the ensuing
15 years, the source has been bright enough for phase self-calibration,
ensuring that the interferometric visibilities are largely free of
the errors and systematic uncertainties associated with poor
atmospheric phase stability. 

\begin{figure}[b!]
%\centerline{\psfig{file=fig_shell.eps,width=0.9\textwidth,angle=270}} 
\centerline{\psfig{file=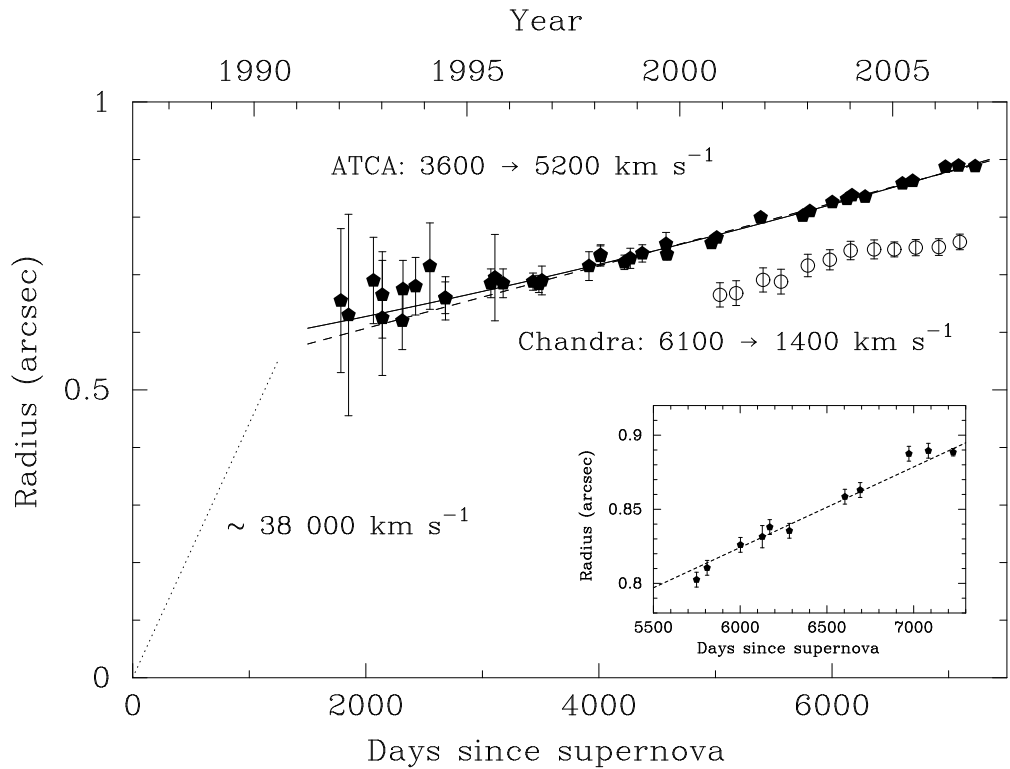,width=0.9\textwidth}}
\caption{The expansion of the remnant of SN~1987A. The filled symbols
plot the radii inferred from thin spherical shell fits to epochs
of 9~GHz $u-v$ ATCA data, while the open symbols plot radii of the
shell seen by the {\em Chandra X-ray Observatory}, as reported by
\cite{pbg+07}. The dashed line is a linear fit to the radio data,
with a slope of $4700\pm100$~\kms. The solid line shows a quadratic
fit, with a slope increasing from $\approx3600$~\kms\ at day 1800
to $\approx5200$~\kms\ at day 7200. The dotted line suggests the
ballistic expansion of ejecta during the initial period when the
remnant was not visible. The inset shows ATCA data over the last
2000 days, with the same linear fit as in the main panel.}
\label{fig_shell}
\end{figure}

The most robust way to track the expansion of the source as a
function of time is not to generate images of the sky, but to fit
directly to the correlated data in the $u-v$ plane
\cite{smk+93}. In particular, \cite{sbr+93} and \cite{gms+97}
showed that the expansion velocity of the radio remnant could be
quantified by fitting the Fourier transform of a transparent thin spherical
shell to each observation.
The results of this analysis, as applied to data through
to the end of 2006, are shown in Figure~\ref{fig_shell}. At the
earliest epochs in this plot, at day $\sim1800$, the diameter of
the shell in the model fit was $\approx1\farcs3$. Assuming that
this source had expanded from size zero at time zero, this implies a
mean expansion speed over the period 1987 to 1990--1991 
of $\sim38\,000$~\kms\
(here and below we assume a distance to SN~1987A of 50~kpc),
consistent with essentially free expansion since shock break-out in
February 1987.

From day 1800 to day 7200, the source size can be fit well (reduced
$\chi^2$ of 0.74) by constant expansion with a velocity of
$4700\pm100$~\kms, shown as a dashed line in Figure~\ref{fig_shell}.
Clearly this implies a dramatic deceleration of the ejecta at around
the time that radio emission was re-detected in 1990, confirming that the
re-emergence of the radio source was due to the shock running into
a new high-density zone in the CSM \cite{cd95,dbk95}.

\begin{figure}[b!]
\centerline{\psfig{file=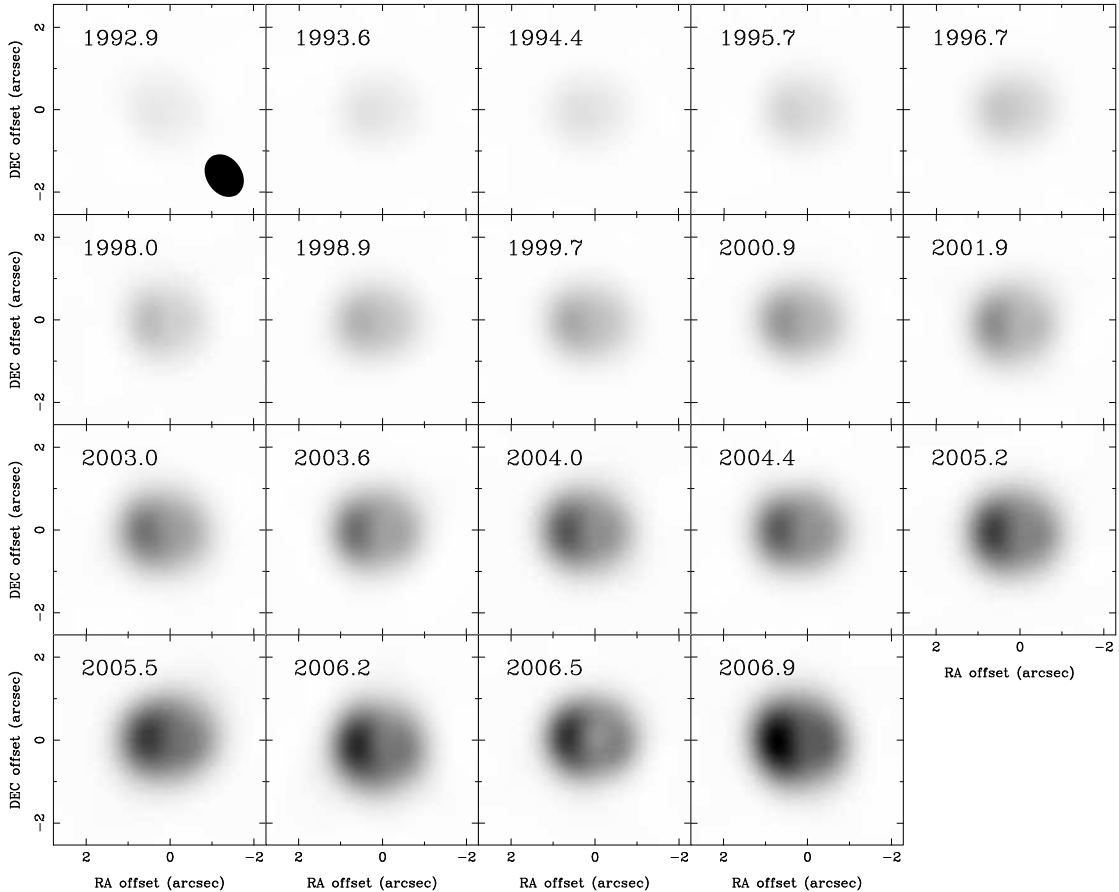,width=\textwidth,angle=270}} 
\caption{9 GHz diffraction limited images of SN~1987A over 14 years of ATCA
observations. Each panel is labeled with its corresponding epoch; the same
linear greyscale range, between --0.2 and +30~mJy~beam$^{-1}$, is used in
each panel. The synthesized beam varies slightly between epochs, but typically 
is an elliptical gaussian of FWHM $0\farcs9$ (the beam for the first epoch 
is shown as an ellipse in that panel).}
\label{fig_natural}
\end{figure}

A slightly better fit to the radio radii is obtained if an
acceleration is included, as shown by the quadratic fit (solid line)
in Figure~\ref{fig_shell}, which has a reduced $\chi^2$ of 0.65.
For this curve, the expansion velocity increases from
$\approx3600$~\kms\ in 1992 to $\approx5200$~\kms\ at the end of
2006.  Conversely, the inset to Figure~\ref{fig_shell}
demonstrates that the last three data-points (covering days 6950
to 7250) suggest a possible {\em deceleration}: a linear fit to
these three observations alone gives a velocity
of $200\pm1800$~\kms, which is
$\approx$2.5$\sigma$ slower than the expansion rate at preceding
epochs.  In these most recent observations, the radio shell has now
reached the same diameter as the equatorial circumstellar ring seen
in optical images.

All of this is sharply discrepant to what has been reported for
{\em Chandra X-ray Observatory}\ data on SN~1987A, taken at comparable
angular resolution. The radius of the X-ray shell as reported by
\cite{pbg+02,pzb+04,pzb+06} is generally $\sim15\%$ smaller than
what we obtain by fitting to the radio data at similar epochs, while
the corresponding X-ray expansion speed was reported to be 6100~\kms\
before day 6100, dropping to $\sim1400$~\kms\ at later times
\cite{pbg+07}.

\section{Imaging, Super-Resolution and Spatial Modeling}

\begin{figure}[b!]
\centerline{\psfig{file=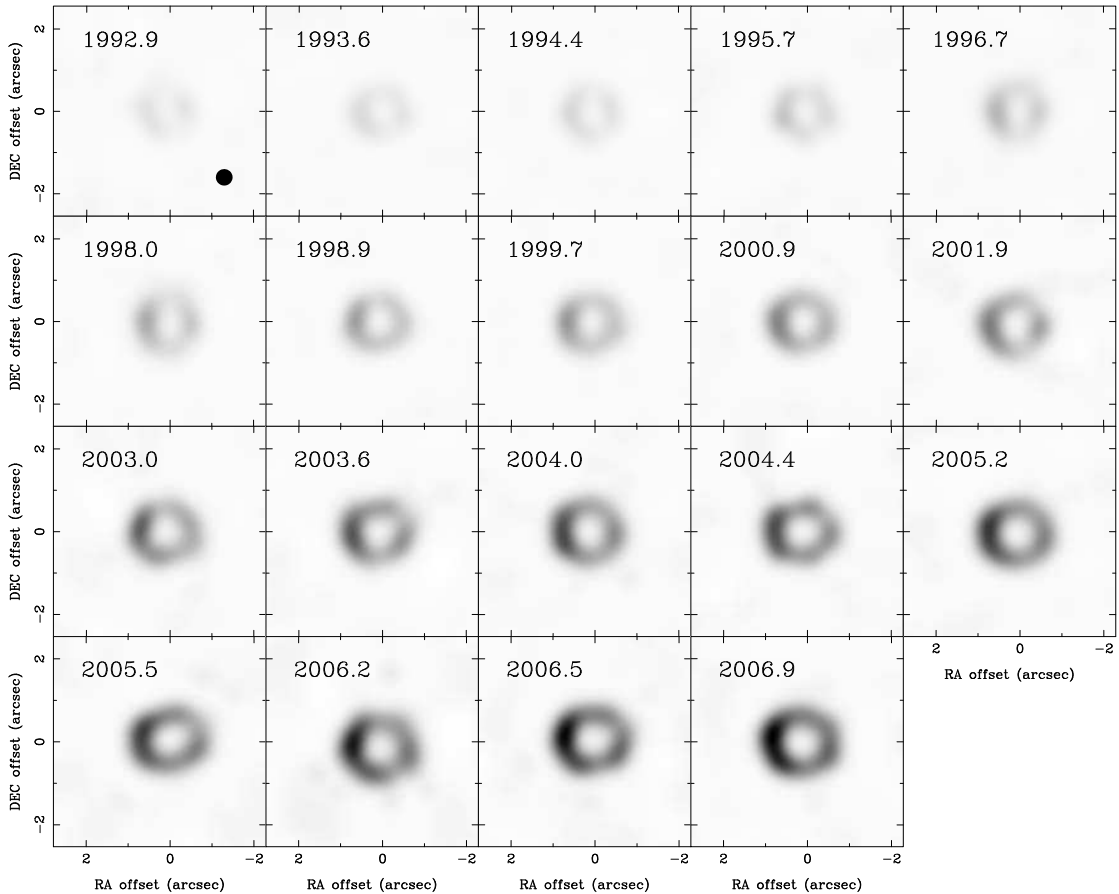,width=\textwidth,angle=270}}
\caption{As in Fig.~\ref{fig_natural}, but after super-resolution has been
applied to data at each epoch. The greyscale now ranges from --0.2 to
+9.8~mJy~beam$^{-1}$, and the synthesized beam for all epochs is a circular
gaussian of FWHM $0\farcs4$ (as indicated in the first panel).}
\label{fig_super}
\end{figure}

Although fits to the $u-v$ data give robust measurements and
uncertainties, they are less well-suited to dealing with complex
morphologies. To move beyond simple spherical shells, we have thus
also imaged the 9 GHz ATCA at each epoch.  While the diffraction-limited
data shown in Figure~\ref{fig_natural} provide only crude morphological
information, application of super-resolution (see \cite{bri94}),
as shown in Figure~\ref{fig_super}, reveals an approximately circular,
limb-brightened shell, dominated by bright lobes to the east and
west. The eastern lobe is brighter than the western lobe, and has
been brightening more rapidly.

\begin{figure}[b!]
\centerline{\psfig{file=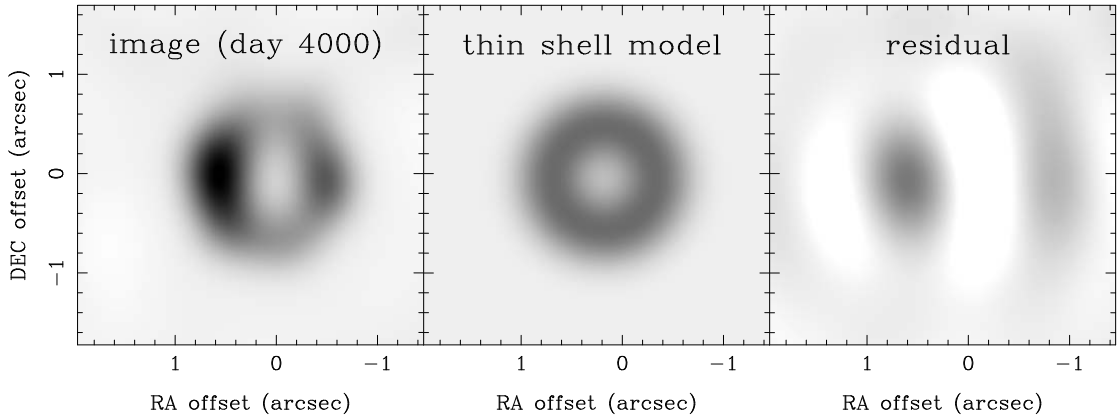,width=\textwidth,angle=270}}
\caption{A fit of a thin spherical shell to the ATCA 9~GHz image of SN~1987A
from epoch 1998.0 (see \cite{mgw+02} for details). The left panel shows
the super-resolved image; the middle panel shows a simulated super-resolved
image generated from the complex visibilities of 
the best-fit thin spherical shell at this epoch;
the right panel shows a super-resolved image of the residual
visibilities between the data and the model. All three panels are shown on a
linear greyscale, 
ranging between --0.2 and +3.0~mJy~beam$^{-1}$ at a resolution of $0\farcs4$.}
\label{fig_1998}
\end{figure}

A more quantitative analysis requires a return to the $u-v$ plane.
As shown in Figure~\ref{fig_1998}, the thin spherical shell analysis
applied in Figure~\ref{fig_shell} is not an ideal fit to the data. It is thus
reasonable to explore the possibility that other simple models might
provide a better fit.

\begin{figure}[h!]
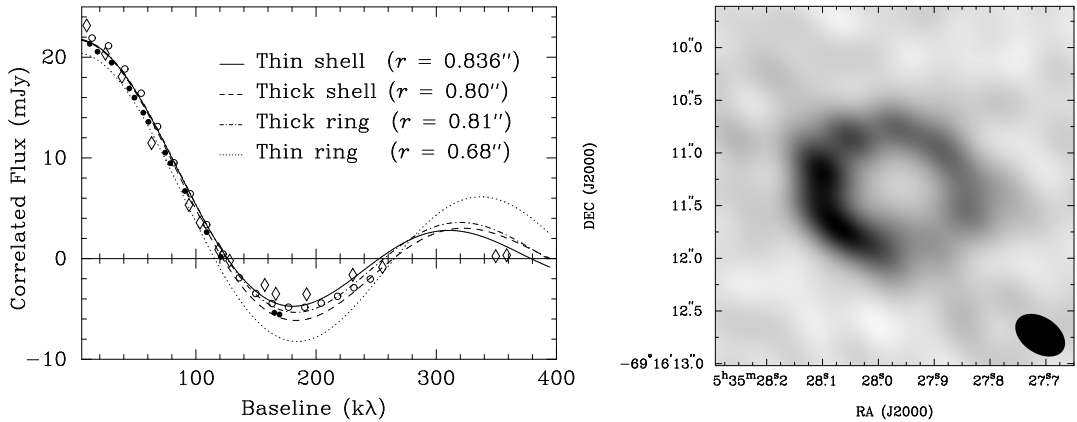

\centerline{\psfig{file=fig_models.eps,height=5.5cm,angle=270}
\psfig{file=fig_12mm.eps,height=5.5cm,angle=270}}
\caption{{\bf Left panel:} Four different fits to azimuthally
averaged $u-v$ data of SN~1987A taken in 2004.  The observations shown are 9~GHz
ATCA data (solid circles), 1.2--8.0~keV {\em Chandra}\ ACIS-S data
(open circles) and 18~GHz ATCA data (diamonds). The flux scale
corresponds to the 18~GHz ATCA data; the amplitudes of the other
two data-sets have been scaled to match this. In all cases the
errors in flux are smaller than the plotted points.  The curves
show the best fits to the 18~GHz data for four different simple
models. The best-fit radius, $r$, is indicated in each case. For
the thick shell and thick ring, the value shown for $r$ corresponds
to the outer radius.  {\bf Right panel:} Diffraction-limited
18~GHz image of SN~1987A, as seen by the ATCA at epoch 2004.4. The
angular resolution is $0\farcs5 \times 0\farcs3$ (as indicated by
the ellipse at lower right) and the greyscale is linear over the
range --0.5 to +3.1~mJy~beam$^{-1}$.}
\label{fig_models}
\end{figure}

A comparison of four spherically symmetric models is shown in the
left panel of Figure~\ref{fig_models}, where we consider fits to
the $u-v$ data of a thin spherical shell, a thick spherical shell,
a thin circular face-on ring and a thick circular face-on ring. If
we try to fit each of these models to the 9-GHz ATCA data, the
angular resolution of which is set by a maximum projected baseline
of $\approx170\,000$ wavelengths, we can rule out a thin face-on
ring, but a thin shell, thick shell and thick face-on ring all match
the measurements well, and imply comparable radii of $\approx0\farcs8$.

Higher resolution spatial information is available from new ATCA
observations at 18~GHz, as reported for epoch 2003.6 by \cite{mgs+05},
and as shown for epoch 2004.4 in the right panel of
Figure~\ref{fig_models}.  These high-frequency data now extend the
coverage of interferometer baselines out to $\approx360\,000$
wavelengths, corresponding to a diffraction limited angular resolution
of $\approx0\farcs5$. Even for these long baselines, the $u-v$ fits
to a thin shell, thick shell or thick face-on ring all fit the data
almost equally well, agreeing to within 5\%.  We thus conclude that
amongst spherically symmetric spatial models, a thin shell is indeed
a reasonable description of the source morphology.

%\begin{figure}[h!]
%\centerline{\psfig{file=fig_12mm.eps,width=0.5\textwidth,angle=270}}
%\caption{Diffraction-limited 18~GHz image of SN~1987A, as seen by 
%the ATCA at epoch 2004.4. The angular resolution is $0\farcs5 \times
%0\farcs3$ (as indicated by the ellipse at lower right)
%and the greyscale is linear over the range --0.5 to
%+3.1~mJy~beam$^{-1}$.}
%\label{fig_12mm}
%\end{figure}

In an attempt to understand the apparent discrepancy between the
reported radii of the radio and X-ray shells (see Fig.~\ref{fig_shell}),
we can Fourier transform {\em Chandra}\ data from a comparable
epoch to provide equivalent sampling of spatial frequencies out to
$\approx220\,000$~wavelengths (mid-way between the 9 and 18~GHz
ATCA data-sets).  As shown in the
left panel of Figure~\ref{fig_models}, the $u-v$
data corresponding to the radio and X-ray observations match to
better than 1\%, indicating that the published mismatch in source
sizes was most likely a result of contrasting measuring techniques,
rather than any physical difference.

Interestingly, the dramatic deceleration of the X-ray expansion at
around day 6100 reported by \cite{pbg+07} is not seen in the radio
data (although, as noted above, deceleration of the radio shell may
have begun to occur at day $\sim7000$).  A possible explanation is
suggested by the observed decomposition of the X-ray image (e.g.,
\cite{pzb+04}: the soft X-ray image closely traces the optical
hot spots seen in the equatorial ring, but in hard X-rays the source
shows two opposing lobes that more closely match the radio data
(see Fig.~\ref{fig_multi}).  The former likely represents
relatively slow moving material interacting with dense gas in the
equatorial circumstellar ring, while the latter corresponds to
higher-velocity ejecta, downstream of the reverse shock, that is
yet to experience significant deceleration.

In this case, we interpret the rate of expansion seen for the radio
shell as the velocity of the reverse shock before the encounter
with the dense equatorial ring.  The range of velocities observed
in Figure~\ref{fig_shell} (3600--5200~\kms) agrees well with the
predictions for this shock velocity of 4100~\kms\ by \cite{bbm97},
and $3700\pm900$~\kms\ by \cite{mmc+03}. The observed acceleration
may be due to increasing amounts of relatively slowly moving SN
ejecta catching up with the shock after its initial sudden deceleration
in or before 1990. On the other hand, since we are now at the point
where the fitted radio radius matches that seen for the surrounding
equatorial ring, we now expect to see deceleration in the radio
expansion, as may already be beginning to occur.

\section{Multi-Wavelength Comparisons}

\begin{figure}
\centerline{\psfig{file=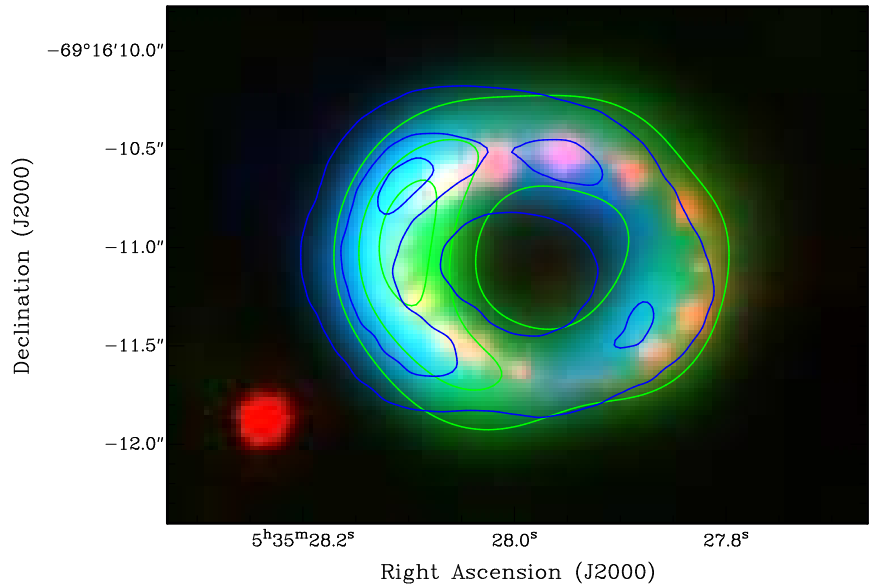,width=0.8\textwidth,angle=270,clip=}}
%\centerline{\psfig{file=fig_multi2.eps,width=0.8\textwidth,angle=270,clip=}}
\caption{A multi-wavelength view of SN~1987A. The observations shown
are an {\em HST}\ WFPC2 F656N image from May~2002 (red image),
1.2--8.0~keV {\em Chandra}\ ACIS-S data from July~2005 (green image and
green contours) 
and a 9~GHz ATCA observation from July~2006 (blue image and blue contours).
The {\em HST}\ and ATCA data are tied to the International Celestial
Reference System (see \cite{rjs+95}), while the astrometry of the
X-ray image is that provided as part of the standard {\em Chandra}\
data products.}
\label{fig_multi}
\end{figure}

A comparison between an {\em HST}\ image of the optical
ring, a {\em Chandra}\ hard X-ray image of shocked gas, and an ATCA radio
image of the expanding synchrotron shell
is shown in Figure~\ref{fig_multi}. The same two-lobed
structure, with brighter regions to the east and west, is seen in
both radio and in hard X-rays; a similar morphology can be
reconstructed from Ly$\alpha$ and H$\alpha$ spectra of the reverse
shock \citep[]{mmc+03,mmp+98,hmz+06}.

\begin{figure}
\centerline{\psfig{file=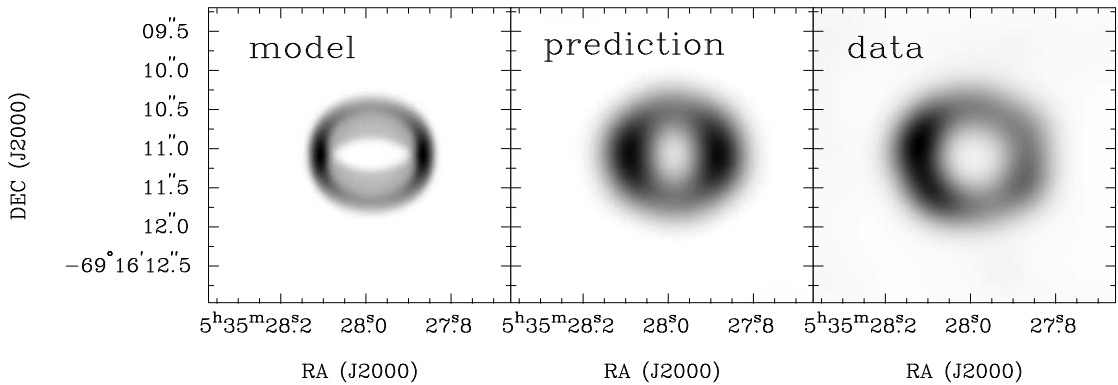,width=\textwidth,angle=270,clip=}}
\caption{Comparison of a tilted ring model to radio observations of SN~1987A.
The left panel shows an optically thin tilted ring, viewed
at an inclination angle of $43^\circ$, and with a thickness  
equal to 25\% of its
radius. The middle panel shows a simulated ATCA observation of this ring,
super-resolved to a resolution of $0\farcs4$.  The right panel
shows a super-resolved 9~GHz ATCA observation of SN~1987A at epoch 2006.9.}
\label{fig_ring}
\end{figure}

It seems clear that the two radio lobes are not related to the
optical hot spots.
The latter are not clustered on the eastern and western parts of
the optical ring, and began to appear many years after the radio
lobes had already been well-established (see detailed discussion
by \cite{mgw+02}).  The one clear match between the optical and radio
data is that the lobes lie along the major axis of the optical ring,
suggesting that the true three-dimensional morphology of the
radio-emitting region can be approximated as a tilted ring or torus,
as shown in Figure~\ref{fig_ring}.  This is consistent with the
expected distribution of the CSM for Sk~--69$^\circ$202, as discussed
earlier: even if the shock driven by the ejecta is close to spherical,
the interaction with the CSM should be occurring primarily in the equatorial
plane, where the RSG wind is densest.  The alignment between the
radio lobe orientations  and the major axis of the optical ring
confirms that the radio emission is indeed dominated by material
emitting in the equatorial zones.

Figure~\ref{fig_ring} also shows what the ATCA would observe for this model,
and compares this prediction
to the actual data. Apart from the brightness asymmetry between
east and west, the match is reasonable. A proper, quantitative
comparison in the $u-v$ plane requires an analytic expression for
the two-dimensional Fourier transform of an optically thin, tilted,
three-dimensional ring, and also needs to incorporate the effects
of light travel times. We are now in the process of undertaking
this more complex analysis.

As noted above, the symmetry is broken by the brightness difference
between the two radio lobes. In both the radio and X-ray bands, 
Figure~\ref{fig_multi} shows that the
brighter eastern lobe is at a larger distance from the explosion
site than the western lobe, suggesting an asymmetry in the initial
velocity of the ejecta, rather than in the ambient density
\cite{gms+97}.

\section{Conclusions}

The turn-on of radio emission from SN~1987A in 1990 and the
accompanying sudden drop in the expansion velocity of
the ejecta marked the initial encounter between the outward moving
supernova debris and the dense red supergiant wind. Over the ensuing
years, the radio remnant of SN~1987A has continued to brighten
and expand.

A joint analysis of radio and X-ray observations of this source
leads to a consistent multi-wavelength picture, in which the reverse
shock is the source of both the synchrotron emission seen in the
radio band, and of the high-temperature shocked gas seen in hard
X-rays. The morphology of this region can be reasonably approximated
by a thin spherical shell, currently expanding at
$\sim$5200~\kms. The symmetry is broken by the presence of two
bright lobes on the eastern and western rims of the radio and X-ray
shells, indicating that the shock is predominantly interacting with
dense gas in the equatorial plane of the progenitor system.  In the
near future, we hope to better quantify the geometry of this
interaction via Fourier modeling of axisymmetric as well as
spherical morphologies.

Clearly the collision of the ejecta piston with dense gas in
the optical circumstellar ring is now almost fully upon us.  As we
watch this spectacular interaction play out over the coming decades,
we can look forward to wider frequency coverage and higher 
resolution images of this remarkable source using forthcoming
facilities such as ALMA, MIRANdA and the Square Kilometre Array.

\begin{theacknowledgments}
We thank the organizers and hosts for a very informative and enjoyable
workshop, and also
thank Sangwook Park and Judy Racusin for providing X-ray data
on SN~1987A in advance of publication.
The Australia Telescope is funded by the Commonwealth
of Australia for operation as a National Facility managed by CSIRO.
B.M.G. and R.N.M. acknowledge the support of the Australian Research Council.
\end{theacknowledgments}

%\bibliographystyle{aipproc}   % if natbib is available
%\bibliography{journals,modrefs,psrrefs,crossrefs}

\end{document}